\documentclass[manuscript]{aastex}

\shorttitle{The WASP Project and SuperWASP Cameras}
\shortauthors{Pollacco et al.}

\begin{document}


\title{The WASP Project and the SuperWASP Cameras}


\author{D.L. Pollacco \altaffilmark{7},
I. Skillen \altaffilmark{3},
A. Collier Cameron \altaffilmark{8},
D.J. Christian \altaffilmark{7},
C. Hellier \altaffilmark{4},
J. Irwin \altaffilmark{1},
T.A. Lister \altaffilmark{8,4},
R.A. Street \altaffilmark{7},
R.G West \altaffilmark{5},
D. Anderson \altaffilmark{4},
W.I. Clarkson\altaffilmark{6},
H. Deeg \altaffilmark{2},
B. Enoch \altaffilmark{6},
A. Evans \altaffilmark{4},
A. Fitzsimmons \altaffilmark{7},
C.A. Haswell \altaffilmark{6},
S. Hodgkin \altaffilmark{1},
K. Horne \altaffilmark{8},
S.R. Kane \altaffilmark{8},
F.P. Keenan \altaffilmark{7},
P.F.L. Maxted \altaffilmark{4},
A.J. Norton \altaffilmark{6},
J. Osborne \altaffilmark{5},
N.R.Parley  \altaffilmark{6},
R.S.I. Ryans \altaffilmark{7},
B. Smalley \altaffilmark{4},
P.J. Wheatley \altaffilmark{5,9}
D.M. Wilson \altaffilmark{4}
           }

\altaffiltext{1}{The Wide Field Survey Unit, Institute of Astronomy,
Madingley Road, Cambridge, CB3 0HA, UK}

\altaffiltext{2}{Instituto de Astrof\'isica de Canarias, C/V\'ia
L\'actea, s/n, E-38200 La Laguna, Tenerife, Spain}

\altaffiltext{3}{Isaac Newton Group of Telescopes, Apartado de Correos
321, E-38700 Santa Cruz de La Palma, Tenerife, Spain}

\altaffiltext{4}{Astrophysics Group, Keele University, Keele,
Staffordshire, ST5 5BG, UK}

\altaffiltext{5}{Department of Physics and Astronomy, University of
Leicester, Leicester, LE1 7RH, UK}

\altaffiltext{6}{Department of Physics and Astronomy, The Open
University, Walton Hall, Milton Keynes, MK7 6AA, UK}

\altaffiltext{7}{Department of Physics and Astronomy, Queen's University
of Belfast, University Road, Belfast BT7 1NN, UK}

\altaffiltext{8}{School of Physics and Astronomy, University of St
Andrews, North Haugh, St Andrews, KY16 9SS, UK}

\altaffiltext{9}{Department of Physics, University of Warwick, Coventry
CV4 7AL, UK}

\begin{abstract}
The SuperWASP Cameras are wide-field imaging systems sited at the
Observatorio del Roque de los Muchachos on the island of La Palma in
the Canary Islands, and the Sutherland Station of the South African
Astronomical Observatory.  Each instrument has a field of view of some
$~$482 square degrees with an angular scale of 13.7 arcsec per pixel,
and is capable of delivering photometry with accuracy better than 1\%
for objects having $V \sim 7.0 - 11.5$. Lower quality data for objects
brighter than $V \sim 15.0$ are stored in the project archive. The
systems, while designed to monitor fields with high cadence, are
capable of surveying the entire visible sky every 40
minutes. Depending on the observational strategy, the data rate can be
up to 100\,GB per night. We have produced a robust, largely
automatic reduction pipeline and advanced archive which are used to
serve the data products to the consortium members.  The main science
aim of these systems is to search for bright transiting exo-planets
systems suitable for spectroscopic followup observations. The first 6
month season of SuperWASP-North observations produced lightcurves of
$\sim$6.7 million objects with 12.9 billion data points.
\end{abstract}


\section{Introduction}

In recent years, interest has grown in relatively small aperture and
inexpensive wide-field imaging systems, essentially composed of large
CCDs mounted directly to high-quality wide-angle camera optics.  The
first prominent success of such an instrument was the spectacular
discovery of the neutral sodium tail of comet Hale-Bopp
\citep{cremonese}, with a temporary purpose-built camera system
(CoCam).  Since then, similar cameras have resulted in imaging of a
gamma-ray burst during the burst period \citep{akerlof}, and the first
detection of the transits of an extra-solar planet in front of its
parent star, HD\,209458 \citep{charbonneau}.  Such instruments are
ideal for projects requiring photometry of bright but rare objects
\citep[see ][]{pinfield05}.

\subsection{Extra-solar planetary transits}

The first extra-solar (exo-) planets were discovered in 1992 by pulsar
timing experiments \citep{wolszczan}. Whilst this technique is
sensitive to the detection of terrestrial-sized planets, its limited
applicability has restricted its use to just a few
objects. \citet{mayor} discovered the first exo-planet, 51\,Peg, from
optical radial velocity studies, and since that time the field has
been dominated by this technique. One of the surprises of these
surveys \citep[e.g. ][]{marcy05} is the existence of a significant
population of solar-type stars accompanied by relatively rapidly
orbiting Jupiter-sized planets.  However, spectral measurements alone
cannot determine unambiguously the true mass or radius of the planet
as the orbital inclination is unknown.  These surveys have discovered
Jupiter-sized objects in orbits out to 3\,AU around 6\% of the nearby
Sun-like stars surveyed.  Of these, some 30\% are {\it Hot Jupiters}
situated in $\sim$4-day $0.05$\,AU orbits, where the equilibrium
temperature is $\sim$ 1500\,K.  About 10\% of Hot Jupiters in randomly
inclined orbits will transit their host star. Therefore, in random
Galactic fields, roughly 1 in every 1000 solar-type stars should
exhibit transits lasting roughly 2 hours with a period of a few days.

Given that the radius of Jupiter is $\simeq$ 0.1\,R$_{\odot}$, these
transits should result in a dimming of the parent star by $\simeq$
0.01 mag. The first transits were discovered in late 1999
\citep{charbonneau}.  The $V = 7.7$ star HD\,209458 is dimmed by
0.016\,mag for 2 hours every 3.5 days, hence both proving the
existence of the planet detected by a radial velocity search, and
resulting in a precise measurement of its radius, mass and bulk
density.  Because of the multiplexing advantage of imaging, this
technique promises to be the fastest way of detecting exo-planets, and
could over the next few years dictate which candidates are followed up
by radial velocity studies (rather than {\it vice-versa} as at
present).

Initially, groups trying to find transits of exo-planets reported
disappointing results. For example the Vulcan Project
\citep{vulcan2002} searched some 6000 stars, finding only 7 transit
like variables. Followup observations of these showed them all to be
stellar in origin.  More recent surveys have have had more success
with published transit detections reported by the OGLE survey (Udalski
et al., 2002a/b), TrES-1 \citep{alonso}, HD\,189733 \citep{bouchy} and
most recently, XO-1 \citep{mccullough}. Part of the reason for the
apparent lack of transits stems from the difficulty in obtaining
photometry of sufficient numbers of solar and late type main sequence
stars. To increase the numbers of stars sampled wide-field surveys
have often concentrated on low galactic latitude fields. However,
while the number of observable stars is undoubtedly increased, we
suspect the stellar population in such surveys is dominated by more
distant K giants.  \citet{brown2003} showed that the number of {\it
bona fide} exo-planet transits (as opposed to stellar impostors) is
consistent with expected numbers of binary and multiple stellar and
exo-planet systems.

\citet{horne2003} lists some 23 photometric transit projects either in
operation or under construction at that time. While many of these are
pencil-beam surveys from traditional telescopes, a number, no doubt
encouraged by the relative cheapness of the equipment, are employing
novel wide field cameras.

\subsection{The WASP Consortium}

The Wide Angle Search for Planets (WASP) Consortium was established in
2000 by a group of primarily UK-based astronomers with common
scientific interests.  In order to reduce the development cycle time
and be on sky rapidly, we use commercially available hardware and
hence limit development work as much as possible.  Our ethos is
therefore quite distinct from other apparently similar projects
e.g. the HAT Project \citep{bakos}.

The WASP Consortium's first venture was the production of the WASP0
camera. Our experience with these types of systems stems from the
CoCAM series of cameras at the Isaac Newton Group of Telescopes from
1996 -- 1998, one of which was responsible for the discovery of the
so-called Sodium Tail in Comet Hale-Bopp 1995 \citep{cremonese}.
WASP0 is composed completely of commercial parts, and utilizes a Nikon
200\,mm, f2.8 telephoto lens coupled to an Apogee AP10 CCD
detector. WASP0 was used in 2000 and 2001 in La Palma and Kryoneri
(Greece) respectively, and has been shown to easily detect the
extrasolar transit of HD\,209458b, amongst other variables (Kane et
al., 2004, 2005a/b).

On the strength of the WASP0 success the Consortium was able to raise
sufficient funding for a more ambitious project -- the multi-detector
SuperWASP cameras. The limited development required is reflected in
the aggressive project timescale: for the La Palma instrument funding
was approved in March 2002 and first light achieved in November 2003,
while for the South African Astronomical Observatory (hereafter SAAO)
sited instrument funding was secured in April 2004 and first light in
December 2005.

In this paper, we describe the SuperWASP facilities at the
Observatorio del Roque de los Muchachos on La Palma (SuperWASP-N) and
the recently commissioned system at the Sutherland Station of the SAAO
(SuperWASP-S). Along with the hardware and data acquisition system, we
outline the SuperWASP reduction pipeline and archiving system for the
data products.

\section{The Hardware System}

In outline each SuperWASP instrument consists of an equatorial mount
on which up to eight wide-field cameras can be deployed. Each is
housed within a two-roomed enclosure which incorporates a roll-off
roof design. Fig.~\ref{FigEnclosure} and \ref{FigCamera} shows the
enclosure at SuperWASP-N and detector system at SuperWASP-S. For both
facilities all observatory functions are under computer control,
including data taking.

\subsection{The Robotic Mount}

Both systems employ a traditional equatorial fork mount constructed by
Optical Mechanics Inc.  (Iowa, USA; formerly Torus Engineering).  The
mount is manufactured within their Nighthawk Telescope range. When
properly configured, the mounts give a pointing accuracy of 30 arc
seconds rms over the whole sky, and a tracking accuracy of better than
0.01 arc seconds per second. The mounts are easily capable of slewing
at a rate of 10 degrees per second. On site the mounts are attached to
a concrete pier.

For our project we do not have a conventional optical tube assembly but
instead we employ a cradle structure to hold the individual
cameras. The cradle allows limited camera movement in 3 dimensions for
balance and alignment purposes.

\subsection{The Enclosure}

The rapid slewing of the mount and large field of view make a
traditional dome impractical and inefficient, hence we have a custom
roll-off roof structure. In most deployments of this design the roof is
moved on to rails overhanging the building, however, in our design the
space under the rails is utilized as a fully temperature-controlled
control and computer room, with the rails integrated into the
roof. The building itself was constructed by Glendall-Rainford
Products (Cornwall, UK), and is manufactured in laminated fiber-glass
strengthened with wood, making the structure extremely rigid.  The
likely absence of a crane during the building erection meant that the
size of the roof panels was optimized to be liftable by 3 people. The
moving roof is controlled by a hydraulically operated ram and
associated electrics. In the case of SuperWASP-N, to fully retract
the roof takes $\sim$19 seconds, and $\sim$54\,seconds to fully
close. The modular design of the building meant the enclosure could be
prefabricated by the manufacturer and then re-erected on site.

With the enclosure roof fully retracted, objects with declinations
$-20 < \delta < 55$ degrees are visible for the entire period when
their altitude is $>$ 30 degrees. For $\delta > 55$ degrees the
movable roof may obstruct visibility at some hour angles. For the
SuperWASP-S instrument we designed a longer enclosure to give
better southern access.

\subsection{The CCDs}

The SuperWASP CCD cameras were manufactured by Andor Technology
(Belfast, UK) and marketed under the product code DW436.  The CCDs
themselves are manufactured by {\it e2v} and consist of 2048 $\times$  
2048 pixels each of 13.5$\mu$m in size. These devices are back illuminated
with a peak quantum efficiency of $>$90\%. Andor use a five stage
thermoelectric cooler to reach an operating temperature of -75C. At
this temperature the dark current is $\sim$11 e/pix/h -- comparable to
cryogenically cooled devices.  As our exposure times are only 30
seconds we do not require this level of performance, and hence we cool
the devices to -50C at which the dark current is a $\sim$72 e/pix/h.

Andor Technology also provide a 32-bit PCI Controller card that is
used to control all CCD functions. These cards (one per detector)
allow the devices to be read out at mega-pixel rates so that even
after all overheads (e.g. header collection, disk write etc), a
new image can be initiated within $\sim$5 seconds of the commencement
of readout of the previous image. Even at this speed the 16-bit images
have good noise characteristics (gain $\sim$2, read out noise
$\sim$8-10 electrons and linearity better than 1\% for the whole of the
dynamic range). To simplify operations, we have not tuned detectors to
controllers in any way.

The original design of SuperWASP-N conceived of using a renovated
existing enclosure with instrument control occurring from a nearby
building. Hence, our shielded data cables are 15\,m in
length. Exhaustive testing showed that at this length data collection
was reliable and mains pickup rarely seen. The detector power supplies
are stored next to the mount.

\subsection{The Telephoto Lenses}

In common with other similar projects, the SuperWASP cameras use Canon
200\,mm, f/1.8 telephoto lenses. These lenses are amongst the fastest
commercially available and have excellent apochromatic qualities.
Funding constraints dictated that we initially purchased 5 lenses from
a local supplier before this format became obsolete.  We subsequently
used {\it www.ebay.com} to track down the remaining units needed for
both instruments. With the above detectors they give a field size of
$\sim64$ square degrees and an angular scale of 13.7 arcsec per
pixel. In the first year of operations for SuperWASP-N our
observations were unfiltered (white light) with the spectral
transmission effectively defined by the optics, detectors and
atmosphere. Subsequently we have deployed broad band filters at both
facilities which define a passband from 400 -- 700\,nm (see
Fig.~\ref{FigWaveband}).

\subsection{The Data Acquisition Computing Cluster}

The easiest method to accommodate the high data rate from 8 cameras is
via a distributed data acquisition cluster, with each detector
controlled by a dedicated DAS ({\it Data Acquisition System}) PC with
local storage disks. Data taking itself is initiated by a central
machine called the TCS ({\it Telescope Control System}), which also
controls more general observatory functions such as pointing the mount
and roof control. The TCS machine also has serial interfaces to a time
service (supplied from a GPS receiver) and weather station. The DAS machines
synchronize time with the TCS through the Network Time Protocol
daemon.  Overall the relative time on the cluster is accurate to
$<$0.1 second, while the GPS system ensures that the absolute time is
accurate to better than 1 second. As the operation of the TCS is vital
to the running of the instrument, a heart-beat system continually
monitors the machine with any break in communication initiating a
close down of the enclosure.

During the night data are stored locally on each DAS machine. At the
end of observing the data are compressed and moved to a RAID system
ready to be copied to tape (LTO2) for transportation back to the UK
(recently SuperWASP-N has begun sending data back via the Internet).

The weather station is provided by Vaisala (foreground in
Fig.~\ref{FigEnclosure}) and has sensors for internal/external
temperature and humidity, wind direction and strength, precipitation
and pressure. A cloud sensor (IR activated) is also utilized.

\section{Data Acquisition Software}

High-level software control of the entire SuperWASP system (robotic
mount, CCD cameras and roll-off roof) is provided by a modified
version of the commercial Linux software {\em Talon} (now Open
Source), produced by Optical Mechanics Inc. (hereafter OMI) for use
with the {\em Torus} mount.  Extensions to the software include
support for multiple CCD cameras (developed by OMI) and some in-house
modifications to add a command-line interface to supplement the
standard graphical interface.

{\em Talon} supports two modes of operation: one for manual control
with an observer present, using the graphical interface (or the new
command-line interface), and the other for automatic observing, where
a dynamic scheduler takes control of the telescope and performs
observations from a pre-defined queue.

In the first season of operation, the observer was responsible for
taking bias and dark frames, opening and closing the dome, and taking
twilight flat fields, automated using a driver script for the
command-line interface.  Science observations were taken using the
{\em telrun} daemon within {\em Talon}, driven from a Perl script. A
new dynamic scheduler, {\em waspsched}, developed in-house by the
Consortium, and using the command-line interface, has recently been
commissioned.  This has increased observational efficiency by allowing
continuous operation of the equipment (previously a number of delays
were required during observing to synchronize the interactions with
the existing {\em Talon} scheduler, allowing about one 30 second
integration per minute despite the 10 degree per second slew speed of
the mount).  The dynamic scheduler also allows us to intersperse
all-sky survey fields with the exo-planet fields. Support for
alternative observing modes may also be added, in particular the
ability to interrupt the scheduled observing and follow up transient
events (e.g. gamma ray bursts) without user interaction.

Data from the site weather station is fed into the software.  {\em
Talon} has a number of configurable conditions on which a weather
alert is issued, for example, high wind or excessive humidity.  On
triggering a weather alert, the telescope is immediately slewed to a
predefined park position (to avoid collisions with the roof), and the
roof is then closed.  After the alert condition finishes, the software
waits for a short period (typically 20 minutes), and then opens the
roof and continues observing if it is still dark.  Other alerts are
generated by the cloud and lightning sensors.  In the event of failure
to operate the roll-off roof, an alert condition is reached sending a
radio signal to a receiver in a neighboring operator attended
telescope dome.

As Andor Technology provides a Linux-based software development kit
all aspects of the CCD control and data collection are integrated
within {\em Talon}.

\section{The Reduction Pipeline}

The SuperWASP data analysis pipeline employs the same general
techniques described by \cite{kane04} for the prototype WASP0
project. We use the USNO-B1.0 catalog \citep{monet03} as the
photometric input catalog. We carry out aperture photometry at the
positions of all stars in the catalog brighter than a given limiting
magnitude.  This has two important advantages for subsequent data
retrieval and analysis; all photometric measurements are associated
with known objects from the outset, and the aperture for every object
is always centered at a precisely-determined and consistent position
on the CCD.

\subsection{Calibration frames}

Bias frames, thermal dark-current exposures and twilight-sky
flat-field exposures are secured at dusk and dawn on every night of
observation. The pipeline carries out a number of statistical validity
tests on each type of calibration frame, rejecting suspect frames
before constructing master bias, dark and flat-field frames.

Master bias frames and thermal dark-current frames are computed by
taking iteratively sigma-clipped means of the ten to twenty frames of
each type taken on each night.  The master bias frame is subtracted
from all thermal darks, flat-field frames and science frames. Temporal
drifts in the DC bias level are removed using the sigma-clipped mean
counts in the overscan region. The overscan strip is then trimmed off
the bias-subtracted frames. The thermal dark frame is scaled according
to the exposure time and subtracted from all flat-field and science
frames.

The twilight sky flats are exposed automatically in a sequence of
fifteen pre-programmed exposures ranging in duration from 1 to
30\,seconds. They are timed so as to be uniformly exposed to a maximum
of about 28000\,ADU in the frame center. The mount is driven to
slightly different positions on the sky between exposures to
facilitate removal of stellar images when the images are combined. The
flat-fields show a circularly-symmetric vignetting pattern, caused by
a sequence of baffles of similar size within each lens. Gradients in
the sky brightness across each flat-field image are removed by
rotating each image through 180 degrees about the center of the
vignetting pattern, subtracting the rotated image, and performing a
planar least-squares fit to the residuals. The gradient is then
divided out from each flat field exposure.

The sky brightness distribution on the short-exposure flat fields is
slightly distorted by the finite travel time of the CCD camera
shutters, which are of the five-leaved iris type. At each pixel
position, we must determine both the correction $\delta t(x,y)$ to the
exposure time and the normalized flat-field value $N_{\infty}(x,y)$
that would be obtained for an infinite exposure time.  The normalized
counts $N(x,y)$ in an image with exposure time $t_{exp}$ are modified
by the shutter travel correction $\delta t(x,y)$:

$$
N(x,y) = N_{\infty}(x,y)\left(1 + \frac{\delta t(x,y)}{t_{exp}}\right).
$$

At each pixel position $(x,y)$, we determine the combined flat-field
and vignetting map $N_{\infty}(x,y)$ and the shutter time correction
map $\delta t(x,y)$ via an inverse variance weighted linear
least-squares fit to $N$ verses $1/t_{exp}$.  An iterative rejection
loop eliminates outliers, which are usually caused by a stellar image
or cosmic ray falling on the pixel concerned in one or more of the
frames.  We then smooth the map of $\delta t$ using a two-dimensional
spline fit and use this to recover an improved map of

$$
N_{\infty}(x,y)=\frac{N(x,y)}{1+\delta t(x,y)/t_{exp}}
$$

for each individual exposure. We then average these corrected
exposures, again using iterative sigma clipping to eliminate stellar
images in individual frames. The shutter correction is applied to the
science frames as well as to the flat fields. In general, for a 30
second integration the shutter corrections are between +0.02 and -0.01
and with a sigma of 0.006. While flat fields are obtained on a twice
daily basis (weather dependent) we use an exponential weighting scheme
to produce a daily master flat field \citep{cameron06}.

\subsection{Astrometry}

In order to derive adequate astrometry, we must establish a precise
astrometric solution for every CCD image.  The celestial coordinates
of the image center can be established to a precision of a few minutes
of arc from the mount coordinates recorded in the data headers, and
from the known offsets of the individual cameras from the mount
position. Subsets of the TYCHO-2 \citep{hog} and USNO-B1.0 catalogs
are made and retained for every pre-programmed pointing of the mount,
and for every camera. The sub-catalogs cover a slightly larger region
of sky than the images with which they are associated, to allow for
pointing uncertainty.

We use the Starlink {\sc extractor} package, which is derived from
SExtractor \citep{bertin1996}, to create a catalog of the $10^4$ or so
stellar images detected at 4$\sigma$ or greater significance on each
frame. We project the TYCHO-2 sub-catalog on the plane tangent to the
celestial sphere at the nominal coordinates of the field center. We
attempt to recognize star-patterns formed by the brightest 100 stars
in both catalogs, and establish a preliminary plate solution
consisting of a translation, rotation and scaling. Further stars are
then cross-identified, and the solution is refined by solving for the
barrel distortion coefficient and the location of the optical axis on
both the sky and the CCD. The RMS scatter of the {\sc extractor}
positions, relative to the computed image positions of TYCHO-2 stars
on the CCD, is always close to 0.2 pixel.

Once the plate solution is established, the pipeline software creates
a photometric input catalog from the list of all USNO-B1.0 objects
brighter than magnitude $R=15$ (in the USNO system) whose positions
fall within the boundaries of the CCD image. Positional and rough
magnitude matching yields USNO-B1.0 identifications for all but a few
dozen of the objects found by {\sc extractor}. These ``orphan"
objects, some of which are likely to be transient variables or minor
solar-system bodies, are added to the photometric input catalog at
their observed pixel locations. In addition, the positions of the
bright planets are computed and, if they fall within the image area,
they are added to the input catalog.

\subsection{Aperture photometry}

The photometric input catalog gives the precise CCD $(x,y)$
coordinates of up to 210\,500 objects, together with their USNO-B1.0
magnitude estimates. We create an exclusion mask for fitting the sky
background, by flagging all pixels within a magnitude-dependent radius
about every object in the input catalog. A quadratic surface is then
fit to all remaining pixels in an iterative procedure. On the
second iteration, the fit is refined by clipping outliers to remove
cosmic rays and faint stars, and adding their locations to the
exclusion mask.

Gradients and curvature in the sky background illumination are removed
by subtracting the quadratic sky fit from the image.  Images are
rejected if more than 50\% of the pixels are clipped or have too high
a $\chi^2$ value - usually indications of significant cloud effecting
the observations.  Aperture photometry is then performed in three
circular apertures of radius 2.5, 3.5 and 4.5 pixels (these apertures
were selected by inspection of images of known blended and unblended
objects, at this spatial resolution). Since the aperture is centered
on the actual star position, the weights assigned to pixels lying
partially outside the aperture are computed using a Fermi-Dirac-like
function.  This is tuned to drop smoothly from 1.0 half a pixel inside
the aperture boundary to 0.0 half a pixel outside it.  The weights of
these edge pixels are renormalized to ensure that the effective area
of the aperture is $\pi r^{2}$ where $r$ is the aperture radius in
pixel units.

The sky background is computed in an annulus of inner radius 13 pixels
and outer radius 17 pixels, so that the sky annulus has ten times the
area of the 3.5 pixel aperture. Pixels flagged in the exclusion mask
as being occupied by stellar images or cosmic rays are excluded from
the sky background calculation.

For a given star, the ratios of the fluxes in the various apertures
contain information on the point-spread function. We define two
wing-to-core flux ratios: $r_1 =( f_3 - f_1)/f_1$ and $r_2 =( f_3
-f_2)/f_2$ where $f_1, f_2, f_3$ are the flux measurements in each of
the three apertures defined above, respectively. Fig.~\ref{FigBlends}
is a plot of $r_2$ against $r_1$ which reveals that for unblended
stellar images, the two wing-core ratios are related by a constant
scaling factor. Stars whose wing-core flux ratios lie close to the
main locus for unblended stars are flagged as such, while outliers are
flagged as likely blends. The photometric measurements for blended
images are very sensitive to small errors in the astrometric fit.
Their light-curves therefore tend to be substantially noisier than
those of unblended objects.

\subsection{Post-Pipeline Calibration - {\sc ppwasp}}

When the photometric input catalog file is created, each object is
labeled with its airmass and catalog magnitudes. The heliocentric time
is calculated on a per object basis (the heliocentric time varies
significantly over the field of view of the instrument). The photometry
modules add information on the sky background level, aperture radius,
raw instrumental aperture fluxes and their associated variances, and
blending information, outputting the results in FITS binary tables.
These tables are then read into the post-pipeline calibration module
{\sc ppwasp} for reduction from raw instrumental to calibrated standard
magnitudes.

{\sc ppwasp} calibrates and removes four main trends in the
photometry: the effects of primary and secondary extinction, the
instrumental color response and the system zero-point. The nightly
mean primary and secondary extinction coefficients are determined from
an iterative least-squares fit to the variation of raw magnitude with
airmass through the night for a sample of stars with colors defined in
the TYCHO-2 catalog. Bayesian priors are used to stabilize the fits on
nights where the airmass range is insufficient to yield a reliable
least-squares fit.  Stars showing excessive variance (determined by a
maximum-likelihood procedure) about the mean trends are down-weighted
by including the excess variance caused by intrinsic variability in
the inverse-variance weights at each iteration.  A time-dependent
adjustment to the primary extinction coefficient is then computed for
each frame, by determining the mean deviation of the ensemble of
calibration stars in each frame from the nightly extinction trend.

Once the instrumental magnitudes have been corrected to a standard
airmass near the middle of the observed range, a linear equation for
the instrumental color response and zero-point of each camera is used
to transform the instrumental magnitudes to a system defined by the
TYCHO-2 $V_t$ bandpass.  Approximately 100 bright, non-variable stars
are adopted as secondary standards within each field. Their
standardized magnitudes as determined over a few photometric nights
are subsequently used to define the ``WASP V" magnitude system for the
field concerned. This allows the final zero-point correction to be
determined to a precision of one or two thousandths of a magnitude for
every frame on every night. This eliminates biases that would
otherwise arise if one or more of the standards were rendered unusable
by saturation or a cosmic-ray hit.

The calibrations are calculated and applied separately for the
measurements from each of the three apertures applied by the aperture
photometry program.  Once post-processing is complete, the calibrated
fluxes are added to the binary FITS tables ready to be ingested into
the archive.

\section{The Archive}

The SuperWASP archive plays a central role in the efforts of the
Consortium, as it is the only long-term repository for the full set of
the WASP photometric data. The archive interfaces have been
specifically designed to facilitate the distributed, collaborative
mode of working that typifies the efforts of the Consortium.

Conceptually the archive comprises three major classes of data: the
bulk processed photometry, the raw images, and an extensible catalog
which can be augmented with the results of various analyses on the
object lightcurves. Simple command-line tools are made available to
the users of the archive to allow access to each of these three data
classes.

\subsection{Archive server hardware configuration}

The architecture chosen for the archive server is a storage cluster,
i.e.  Beowulf-style cluster comprised of commodity compute nodes with
Gigabit Ethernet inter-connect, wherein each cluster node is fitted
with a large disc capacity (1.2TB per node in the first
incarnation). This model has a number of benefits -- not least in that
it provides the compute power required for large-scale data-mining
activities in a cost-effective manner. Perhaps as important however is
that by embedding and distributing the storage throughout the cluster
all of the nodes act both as data servers and data consumers.  There
are multiple independent network paths between the servers and their
potential consumers. This alleviates potential network bottlenecks
associated with a more traditional model in which data is stored on
and served by a small number of high capacity storage nodes and
consumed by a large number of client nodes. It also adds resiliency,
in that the failure of a single node impairs archive performance for
only as long as it takes to restore the relevant data to a different
node. Lastly this model allows us to scale total capacity simply by
adding more nodes, with each additional node contributing additional
storage, compute power and overall network bi-section bandwidth.

\subsection{Data storage}

Within the archive the raw images and the processed photometric data
are held in the form of FITS files on the archive server. These files
are then indexed using a {\it Relational Database Management System}
(RDBMS) which allows the data relevant to a user query to be found in
a fast and efficient manner.  The RDBMS is also used to store and to
support queries on the WASP catalog. MySQL was chosen ahead of
PostgreSQL during the design phase based on performance measurements
for typical queries anticipated at that time.

An early and important design decision made during the development of
the archive was to separate the image and photometric data from the
catalog and indexing information stored within the tables in the
RDBMS. Whilst in principle it would be possible to store images and
lightcurves as BLOBs (Binary Large OBjects) within the RDBMS itself,
we felt that for the data volumes to which the WASP archive is
expected to scale (30\,TB photometry and $>$100\,TB images over three
years) such an implementation would be risky, with many potential
performance and data integrity issues arising as the collection
accumulated. The choice of FITS as a format to store the data was
motivated by the wide acceptance of this format within astronomy, and
the availability of stable and efficient third-party libraries and
tools (for example CFITSIO) to read and write these files.

The arrangement of the raw image data within the archive is relatively
straightforward -- the image files as processed by the pipeline are
stored in their original form, and are located within a directory
hierarchy on a hierarchical storage management system.  An entry into
the RDBMS allows images matching a user query to be identified
rapidly.

For the photometric data the situation is rather more complex. The
WASP pipeline at its heart operates on a single image at a time; the
output of the pipeline is a collection of files, each of which
contains the calibrated photometry for all of the stars detected in a
given image. A typical delivery of data from the pipeline to the
archive will be a collection of such files representing the results of
processing the data from a single night of observations. The end-user
of the archive will in the overwhelming majority of cases be
interested in obtaining true light-curves of objects of interest, i.e.
a single file containing the photometry of a single object, collated
from a large number of images taken over many weeks or months. For
this reason it is necessary for the archive to re-order the
photometric data before it is presented to an archive user.

To create a lightcurve of an object from the pipeline products it is
necessary to be able to associate each photometric data point with a
unique object; by matching object identifiers from image to image, it
is then straightforward to build up a complete lightcurve. This
process is greatly aided by the catalog-driven nature of the WASP
pipeline -- each data point can be uniquely and unambiguously assigned
an identifier based on its place in the source catalog
(USNO-B1.0). This obviates the need for complex and potentially
error-prone cross-identification between photometric points based on
positional coincidence, for example.

The collation of photometric data points into complete lightcurves
takes place in two stages within the archive. The first stage occurs
during the ingest phase, and involves the re-ordering of the data from
a given night (comprising several thousand files) into a smaller
number of new files in which the photometric points for each object
are stored contiguously within the rows of a FITS binary table. These
files are also indexed internally by object name, which allows the
table rows containing the photometry for a given object to be located
rapidly. To ensure that the size of these bulk storage files is kept
within manageable bounds the archive divides the sky into {\it
sky-tiles}, based on a Plate Carree projection of equatorial
co-ordinates. The size of these files is chosen to be 5$\times$5
degrees, as this will typically limit the size of the storage files to
be $\sim$100\,MB to $\sim$1\,GB.  So the result of this re-ordering
step in the ingest phase is a relatively small number of bulk storage
files, each containing a single night of photometric data for all
stars in a given sky-tile (typically numbering thousands), presented
in object order. These bulk files are then assigned a location on the
storage cluster, and indexed in the RDBMS.

For each sky-tile a meta-index is created by collating the internal
indexes in the per-night files associated with that sky tile. This
meta-index lists, for each cataloged object, which files contain data
pertaining to that object, and the numeric indices of the first and
last rows in the table in each file holding the data. The sky-tile
meta-index is written to disc as a FITS binary table file, however to
improve access times the meta-index files are also copied into a
ram-disk on the archive server cluster nodes.

The second phase of lightcurve collation occurs during the extraction
process in response to a user query. A search of the RDBMS determines
in which sky-tile the object of interest falls. A scan of the
meta-index for that sky-tile then yields the names of the bulk
photometry files containing photometry for the object of interest,
plus the row offsets into the binary table. These files are then
opened in turn, the photometric data points are read and collated in
memory into a FITS format lightcurve which, once complete, is
presented to the user. This three-stage indexing process is highly
efficient; tests on the system have demonstrated that a lightcurve
comprising $\sim$10,000 data points and spanning 130 days can be
extracted in a little as 850 milliseconds. This speed of access is of
critical importance to many of the data-mining activities that will
take place on the WASP archive, some of which will potentially need to
access millions of lightcurves.

The pipeline processing of WASP images also yields detections of
transient and/or uncatalogued objects. These ``orphan'' data-points
are handled separately from the standard catalogue-driven photometry,
and are stored in a table within the RDBMS for later analysis.

\subsection{The catalog}

The second key element of the archive is the WASP catalog. At its core
this provides very basic information about each object observed by
SuperWASP, for example equatorial co-ordinates and its relation to the
entry in the USNO-B1.0 catalog that is used to drive the photometric
pipeline.  The catalog has been designed to be extensible however,
recognizing that it is nigh on impossible to define {\it a priori} the
additional per-object attributes that the catalog might be called upon
to store as the understanding and exploitation of the WASP data
proceeds.

The catalog extensibility is implemented by storing newly derived
attributes in separate tables in the RDBMS, alongside a WASP object
identifier that acts as a key. By making use of table joins within the
RDBMS at query-time the root catalog can appear to the end-user to be
extended on-the-fly to include these new attributes.  The implication
of this is that the results of any data-mining effort, for example
variability testing or period searches, can be loaded into a table in
the RDBMS and immediately become visible to all users of the archive
as additional per-object attributes which can be used in queries in a
transparent manner. In situations where different groups have
different algorithms for attacking a problem (a notable example being
planet-transit searches), these results can be ingested into different
tables, and the choice of which set of results to use for any given
query is left to the end-user, rather than the architect of the
archive.  This flexibility means that the archive will potentially
become a very powerful tool in the exploitation of the WASP dataset,
evolving to meet the needs of the end-users.

Internally to the archive queries of the RDBMS are constructed using
the industry-standard {\it Structured Query Language} (SQL). To
obviate the need for non-expert users of the archive to formulate SQL
queries directly (queries which can become very complex when involving
table joins), we have developed a simpler language in which queries
can be defined as a set of filters on object attributes using a much
simpler syntax than SQL. This language (which we call WQL, the {\it
WASP Query Language}) can be readily compiled into SQL queries by the
archive software.  Amongst other features, WQL allows users to:
perform cone searches by object co-ordinate (including on-the-fly name
resolution using SIMBAD); to filter on object attributes; to select
which attributes will be displayed in the output and to sort the
output on a selected attribute.  The results from the query are
returned either as formatted ASCII tables in a number of
user-selectable flavors, or as HTML. WQL is not directly comparable to
other query languages, for example the Astronomy Data Query Language
(ADQL) which has an SQL-like syntax. While ADQL is undoubtedly more
powerful and flexible than WQL, it suffers in that it is not very
user-friendly. The design goals of WQL were to create a query language
which provides sufficient power and flexibility for most normal use
cases, but does so using a syntax which is amenable to use by
astronomers without a computer science background. As a future
development the WQL compiler may be retargetted to generate ADQL
rather than SQL queries.

\subsection{The user interface}

The primary mechanism by which users will access the WASP archive is
by means of three command-line tools which enable the querying of the
catalog, the extraction of lightcurves, and the extraction of raw
images. These tools communicate with the archive server using the HTTP
protocol. HTTP was chosen as it potentially enables a wide variety of
other clients (in particular web browsers) to be used in the future,
and also because HTTP requests generally pass unmolested through
network firewalls (which would not be guaranteed of a bespoke protocol
on an arbitrary TCP/IP port).

The advantage of HTTP is that every WASP image and lightcurve have a
unique and predictable URL, which can be accessed using any tool
capable of retrieving a file via HTTP. Tools which are capable of
opening FITS files directly from URLs (e.g. any application built on
top of CFITSIO) can therefore request and open a lightcurve or image
from the WASP archive across the Internet with no more difficulty than
if it was on local disc, simply by specifying a URL rather than a
filename. In this respect the WASP archive acts as a web service.

\section{Observatory Performance}

\subsection{Hardware}
The integrity of each observatory is checked remotely on a daily basis
using interior and exterior network cameras, and manually on a weekly
basis as part of a comprehensive preventative maintenance plan.
Furthermore, during routine observing, the operation of the various
subsystems can be monitored remotely using HTTP clients.  No
significant malfunctions have occurred to date. On La Palma, the
enclosure has withstood three harsh winters, which included near
200\,km/h winds in a tropical storm, and 1.5-m of snow, without
adverse effect.  After the first season of operations we found some
indications of wear on the hour angle friction wheel manifesting
itself as vibration while slewing. After some investigation we
realized that during this first season we had some inexperienced
observers that were moving the mount without disengaging the
clamps. For subsequent seasons we have avoided this problem by redoing
the pointing model and using the unused half of the friction wheel.
Other problems arose in the declination axis: the original design of
the mount expected a normal telescopic optical tube assembly, but the
need to be able to insert the SuperWASP cradle meant that some
movement was required. In the original design the declination axis
encoder was held at the other end to the drive. As this encoder is not
a packaged product but a reader and glass disc movement of the axis
meant it was difficult to keep the encoder focused (or worse). For the
second mount we therefore requested OMI move the encoder to the drive
end of the axis. For SuperWASP-N this problem only manifests itself
when we are servicing the cradle unit. In normal operations these
mounts are made to work extremely hard. For SuperWASP-N the first
season of operations regularly led to more than 600 movements per
clear night. Similarly, the thermoelectrically cooled detectors have
proved to be extremely reliable, and operate stably at -50C even when
the ambient nighttime summer temperature is in the mid 20C's.

\subsection{Pipeline}
The pipeline has proved robust and runs in an unattended fashion. For
example, an average night during the 2004 season produced around 550
frames per camera of 16 fields. After the post-processing and data
quality control around 25 were rejected as outliers in the astrometric
or point spread profile fits (e.g. trailed images). From the 16 fields
around 710000 star-like objects were accepted with good certainty and
statistics while a further 220000 were rejected as they had a below
threshold number of data points (either because they were too faint or
that they saturated the detector).  Each frame produced from $\sim$200
orphan objects (mostly cataloged objects in the USNO-B1.0 but with
discrepant brightnesses). We find that on an average Pentium-4 (3GHz
processor, 1Gb memory, SCSI disk system) class machine one night of
data from one camera takes about 11 hours to reduce.

For the 6 months that SuperWASP-N was operational in 2004, the
instrument produced lightcurves for $\sim$6.7 million objects
containing 12.9 billion data points.

\subsection{Archive}
A typical archive catalog query will return results within 1-10
seconds, depending on the complexity of the query and the number of
matches. The RDBMS makes extensive use of in-memory caching of
database tables, so repeating a query with small changes to the
clauses will
 often result in much improved performance for the second
and subsequent queries.

Extraction of an arbitrary lightcurve will typically take $\sim$1
second, however the operating system in the storage cluster nodes will
take the opportunity to cache some of the contents of the bulk
photometry files in memory.  Subsequent queries for lightcurves from
nearby objects on the sky can often return much more quickly than the
initial request. The clustered nature of the archive server allows
dozens of independent requests to be served simultaneously without
noticeable loss in performance.

Initially the WASP archive will remain a private facility, however, if
and when additional resources become available we will make the data
publicly available.


\subsection{Photometric Performance}

While our design goal was to reach 1\% photometry for stars brighter
than $V \sim 12.5$, in practice we reach this level of precision for
stars brighter than $V \sim 11.5$. Stars brighter than $V \sim 9.4$
have a precision better than $0.004$ mag.  Equally important, the
night to night variations are $< 0.002$ mag.  Objects with $V < 15.0$
are stored in the archive. Fig.~\ref{rms2} shows the rms errors as a
function of magnitude over the 2004 season and demonstrate that the
complete system delivers good precision over an extended period. The
faint end of our magnitude range is dictated by the sky brightness: in
dark conditions the integrated sky brightness in one of our pixels is
$\sim$16.5.

\subsection{Example Lightcurves}

For relatively large amplitude variations ($\Delta V > 0.05$) data
from the archive is of sufficient quality for identification of
variable stars (etc). Fig.~\ref{gn_boo_ps} is an example of a large
amplitude eclipsing system extracted from the archive. We stress this
figure is composed completely of raw extracted data and has not
undergone any further processing. In this figure there are 3873 data
points taken over a 6 month period through all prevailing weather
conditions. Close inspection shows that after ignoring outliers the
rms errors are around 0.015 magnitudes at the quadratures where it has
a higher than normal noise for its magnitude.

Our primary design goal has always been to identify transit amplitudes
with $\Delta V \sim 0.01$. In this regime secondary effects can make
the photometry less reliable. Rather than analytically trying to
remove these we use the {\sc SYSREM} algorithm discussed by
\citet{tamuz} to de-trend the data. This algorithm removes time and
position dependent trends from the lightcurves by computing the
weighted average of magnitude residuals over all stars to calculate a
time-dependent variation. We added an additional weighting to the
magnitude uncertainties to down-weight known variable stars and
poorer-quality images. Note the {\sc SYSREM} algorithm does not
identify the causes of the trend, but removes it blindly, using all
the stars available; for this reason it should be applied with caution
where finding genuine photometric variables is the desired
outcome. After some experimentation we remove four trends from the
data. Fig.~\ref{ExampleLC2} is an example of a transit candidate with
and without the {\sc SYSREM} algorithm applied. This candidate is
identified as 1SWASP J133339.36+494321.6 and has $Vt = 11.154$
mag. Out of transit the rms scatter in the lightcurve is 0.0076 mag.
Other exo-planet candidates from the first SuperWASP-N season of 
observations are now becoming available \citep[e.g. ][]{christian06}.
For details of our {\sc SYSREM} implementation and transit candidate
identification algorithm see \citet{cameron06}.


\section{Summary and Future Development}

The original SuperWASP Camera was constructed at Observatory of the
Roque de los Muchachos on la Palma in 2003. Since then we have also
constructed a clone facility at SAAO which has recently seen first
light.  Each system has a total field of view of $~$482 square degrees
and is capable of monitoring the sky down to $\sim 15$th magnitude
every $\sim 40$ minutes. For the La Palma facility, in the first
season of operation the instrument was used solely to monitor fields
with high cadence ($\sim$10 minutes). In subsequent seasons the
development of the dynamic scheduler will allow all sky monitoring
interspersed with the high cadence exo-planet fields (we expect
complete sky coverage once per night).

The reduction pipeline is a continued development from the WASP0
instrument now generalized and expanded to cater for both
facilities. The rms plots show that the precision at the bright end of
our magnitude range reaches 3-4 milli-magnitudes over an extended
period and stars brighter than $\sim$11.5 magnitude have errors less
than 1\%.  An average night's data of 50\,Gb, produces some
6$\times10^6$ records in the archive.

We have developed an archive system capable of storing tens to
hundreds of terabytes of raw and processed data, the results of
high-level analyses of those data, but doing so in a manner that
provides rapid access to individual lightcurves and images. The
central catalog is designed to remain flexible and responsive to
evolving user requirements despite the burgeoning data volume. The
choice of HTTP as an interface protocol allows the end-user a great
deal of flexibility in the choice of access clients, and facilitates
truly distributed, collaborative working within the Consortium.

 From the start of operations in 2006 at both facilities we have begun
an all-sky survey mode. As the exo-planet program will continue we
have integrated these different survey modes (with only a small loss
to the exo-planet survey efficiency) enabling once a night coverage of
the entire visible sky.  To do this effectively we will eventually
deploy a reduction pipeline at the sites themselves. By compromising
performance slightly we will be able to use this to grade the quality
of the night, directly feeding back to the dynamic scheduler in order
to modify the observing queue, hence matching the science to the
prevailing weather conditions. In principle this could also provide a
real time alert system and if sufficient software resources become
available we could use this to arrange followup observations on other
nearby facilities.

The SuperWASP Cameras are now surveying large parts of the sky
primarily in search of bright, transiting exo-planets. In the near
future we will present large numbers of high quality light curves
suitable for followup and expect to at least double the numbers of
known transiting exo-planets. Interests of the Consortium members also
cover other ``secondary'' science areas and publications in these
areas will be forthcoming. For non-members of the consortium we also
offer the opportunity to present ideas to the Project Steering
Committee and hence achieve a Guest Observer status in this area.


\section{Acknowledgements}

The WASP Consortium comprises scientists primarily from the University
of Cambridge (Wide Field Astronomy Unit), the Instituto de
Astrof\'isica de Canarias, the Isaac Newton Group of Telescopes, the
University of Keele, the University of Leicester, The Open University,
Queen's University of Belfast and the University of St Andrews.

The SuperWASP Cameras were constructed and are operated with funds
made available from the Consortium Universities and the UK's Particle
Physics and Astronomy Research Council. We would also like to thank
the UK's STARLINK project (and site managers) for their invaluable
support in the development of the pipeline. 

We thank the Administrador del Observatorio del Roque de los
Muchachos, Sr. Juan Carlos Per\'ez, and the Director of the Isaac
Newton Group of Telescopes, Dr Ren\'e Rutten, and staff, for their
invaluable support in the construction and operation of
SuperWASP-N. We would also like to thank SAAO for hosting SuperWASP-S
and thank SAAO personnel for their generous assistance in the
planning, construction and maintenance of SuperWASP-S.
We would like to acknowledge the anonymous referee who made an 
important contribution to this paper.

This publication makes use of data products from the Two Micron All
Sky Survey, which is a joint project of the University of
Massachusetts and the Infrared Processing and Analysis
Centre/California Institute of Technology, funded by the National
Aeronautics and Space Administration and the National Science
Foundation.

\clearpage



\begin{figure}
  \centering
     \caption{The SuperWASP-N enclosure at the Roque de los Muchachos
    Observatory on la Palma. The weather station is in the background.}
             \label{FigEnclosure}%
\end{figure}

\begin{figure}
  \centering
     \caption{The SuperWASP-S instrument with eight cameras
mounted. The field of view of the instrument is $\sim$482 square
degrees.}
             \label{FigCamera}%
\end{figure}

\begin{figure}
  \centering
     \caption{Passband of the new SuperWASP filter (top panel) plotted
alongside the atmospheric transmission, CCD response and lens
transmission. The bottom panel shows the original unfiltered system
alongside those of the new filter and the Tycho\,2 V filter.}
             \label{FigWaveband}%
\end{figure}

\begin{figure}
  \centering
     \caption{Graph showing that wing-to-core ratios defined from
different sized apertures are related by a constant scaling factor.
This plot does not include all stars in a field. Instead only those
that are brighter than $V = 15.0$ and have reasonable values for the
$r_1$ and $r_2$ indices ($< 0.5$ for both) are included. The pipeline
actually flags 9 different types of ``blended'' object ranging from
type 0 (unblended) to 9 (saturated star or a bad pixel). The types
shown in this figure correspond to extremely red objects (1) or those
indices that suggest nearby companions or distorted images (2-4).
The vertical line shows when the effect of nearby companion stars is
detectable in the $r_1$ index and was derived from real SuperWASP-N data
in conjuction with 2MASS and DSS data.}
            \label{FigBlends}
\end{figure}

\begin{figure}
  \centering
     \caption{An rms error verses magnitude diagram for the 2004
season. Stars brighter than about $V \sim$11.5 reach better than 1\%
accuracy. Also shown is a theoretical model which shows that the noise
is dominated by the sky contribution (uncertainties in the sky
contribution are apparent at fainter magnitudes).}
     \label{rms2}%
\end{figure}

\begin{figure}
  \centering
     \caption{An example of a eclipsing binary star (GN\,Boo) from WASP
data. This object has a period of 0.301601 days. This light curve is
composed of 3873 data points obtained over the 2003 season obtained in
{\it all} weather conditions and has not undergone any post archive
processing.}
     \label{gn_boo_ps}
\end{figure}

\begin{figure}
  \centering
     \caption{A typical transit candidate, 1SWASP J133339.36+494321.6
found in SuperWASP data. The top lightcurve shows the archived data,
while that below shows the same data after detrending through the {\sc
SYSREM} algorithm. This candidate has a transit of amplitude
$\sim$1.5\% and duration 2.8\,hr. The data is phased on a period of
3.63 days. Published data for this object available in the Tycho-2 and
2MASS catalogs suggest the host has a spectral type of F7 (from the
$V_{t(SW)} - K_{2MASS}$ color index; the 2MASS $J-H$ index give a
spectral type of F8). Consequently, the radius of the companion is
1.23\,R$_J$. As these values are derived from photometry they serve as a
guide, but further observations of this target will be necessary to
confirm its status. 
}
      \label{ExampleLC2}%
\end{figure}

\clearpage

\begin{figure}
  \centering
     \includegraphics[width=120mm]{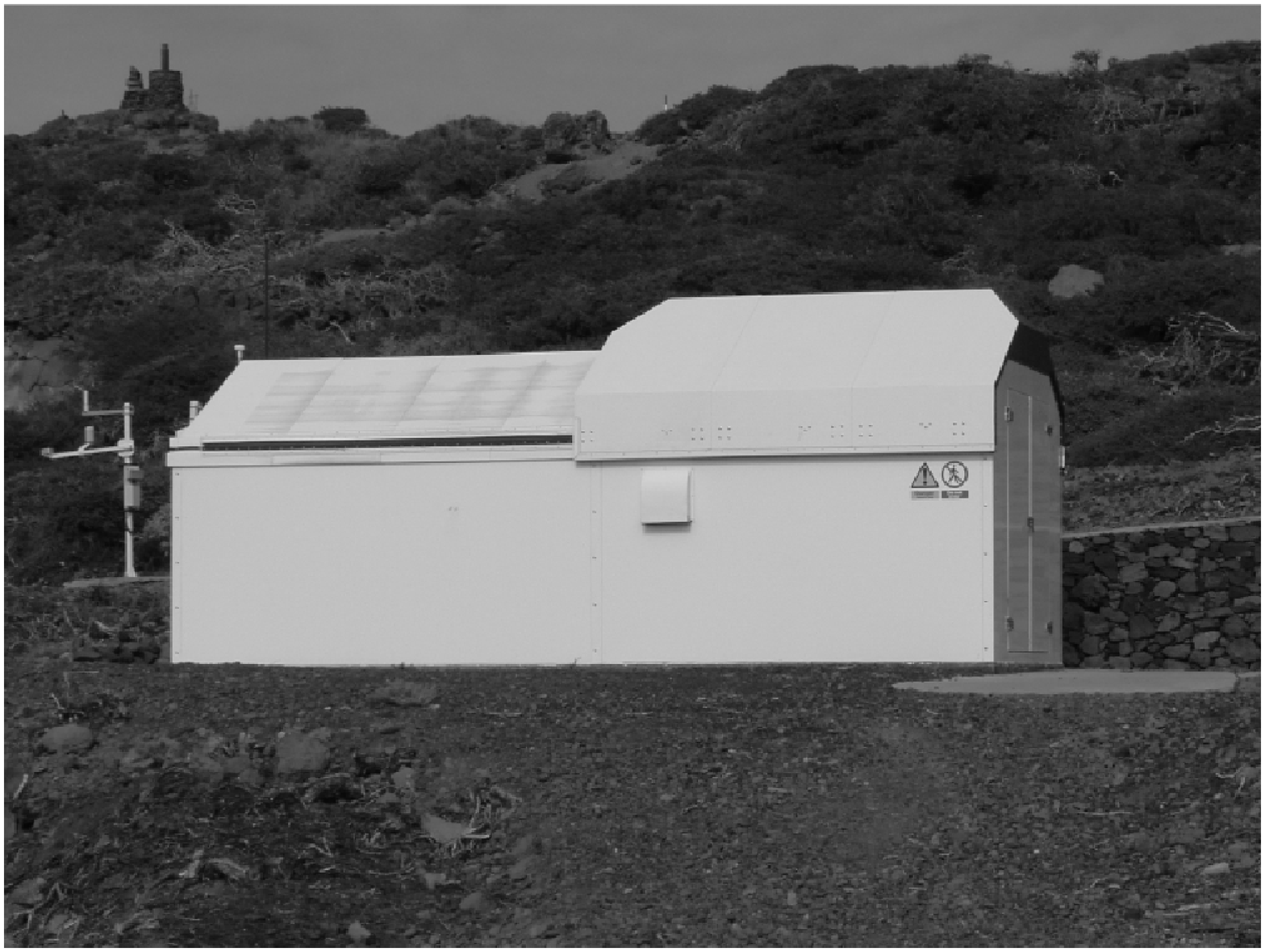}
\end{figure}

\clearpage

\begin{figure}
  \centering
     \includegraphics[width=120mm, angle=-90]{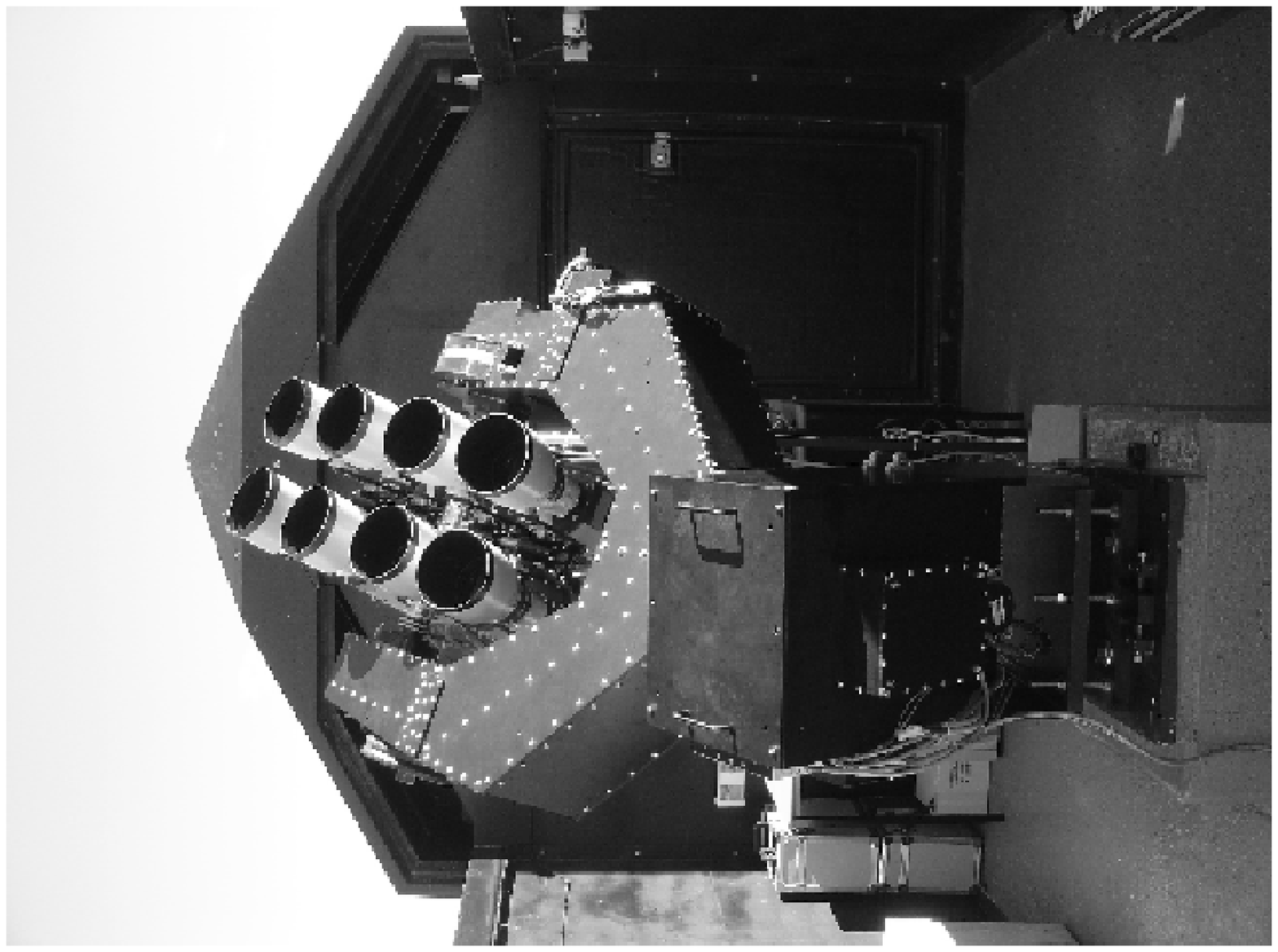}
\end{figure}

\clearpage

\begin{figure}
  \centering
     \includegraphics[width=120mm]{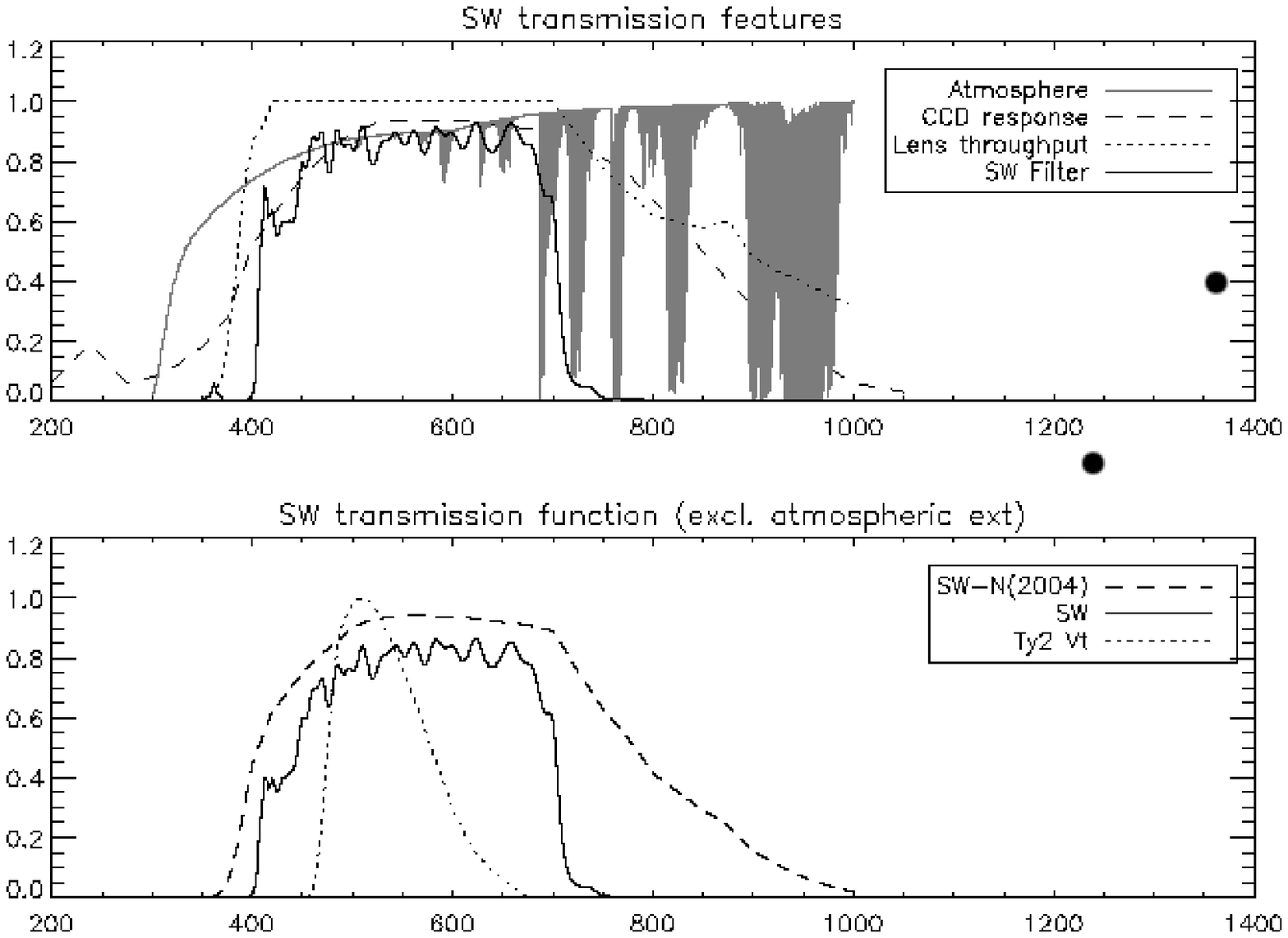}
\end{figure}

\clearpage

\begin{figure}
  \centering
     \includegraphics[width=120mm, angle=-90]{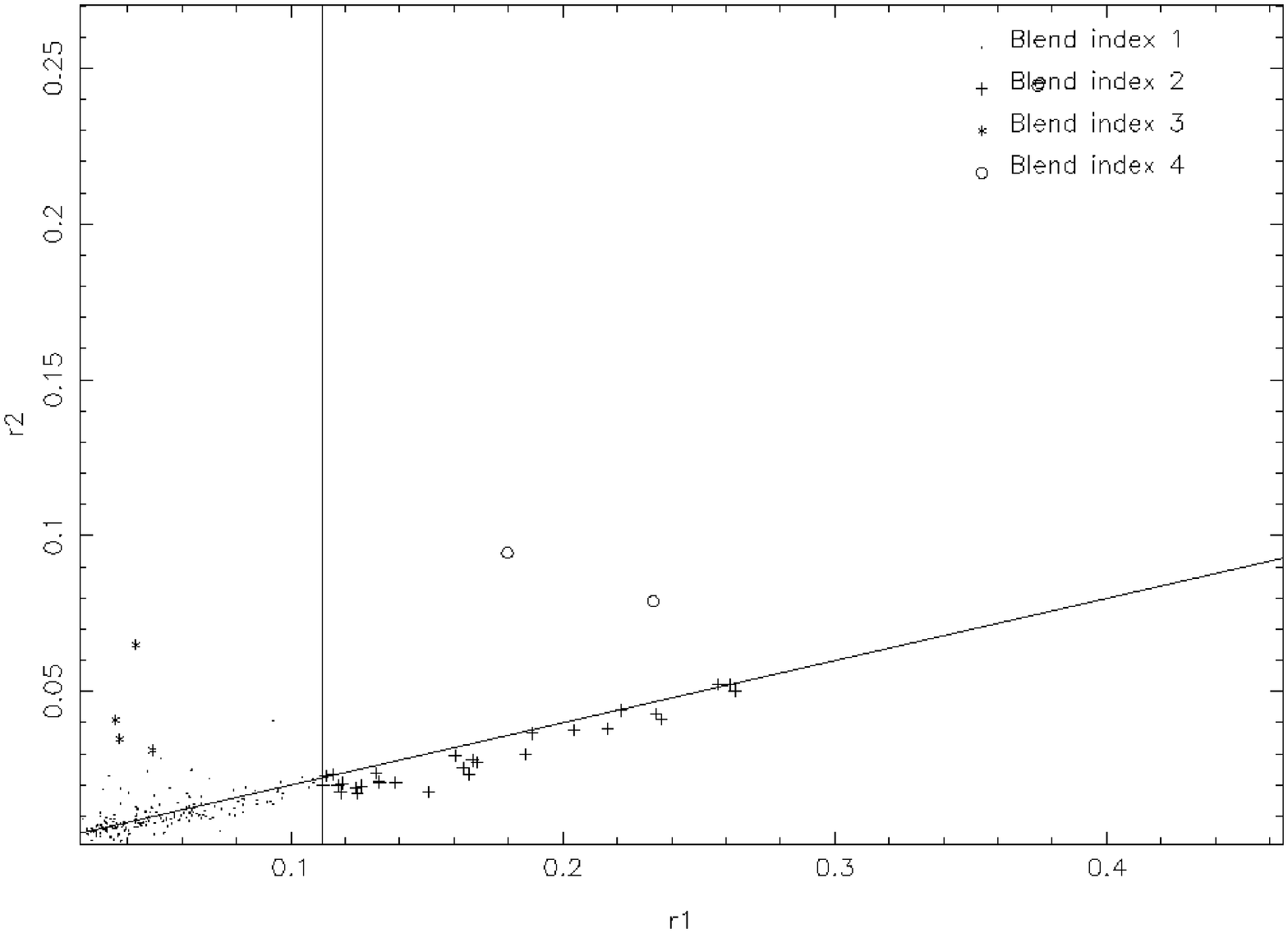}
\end{figure}

\clearpage

\begin{figure}
  \centering
     \includegraphics[width=100mm, angle=-90 ]{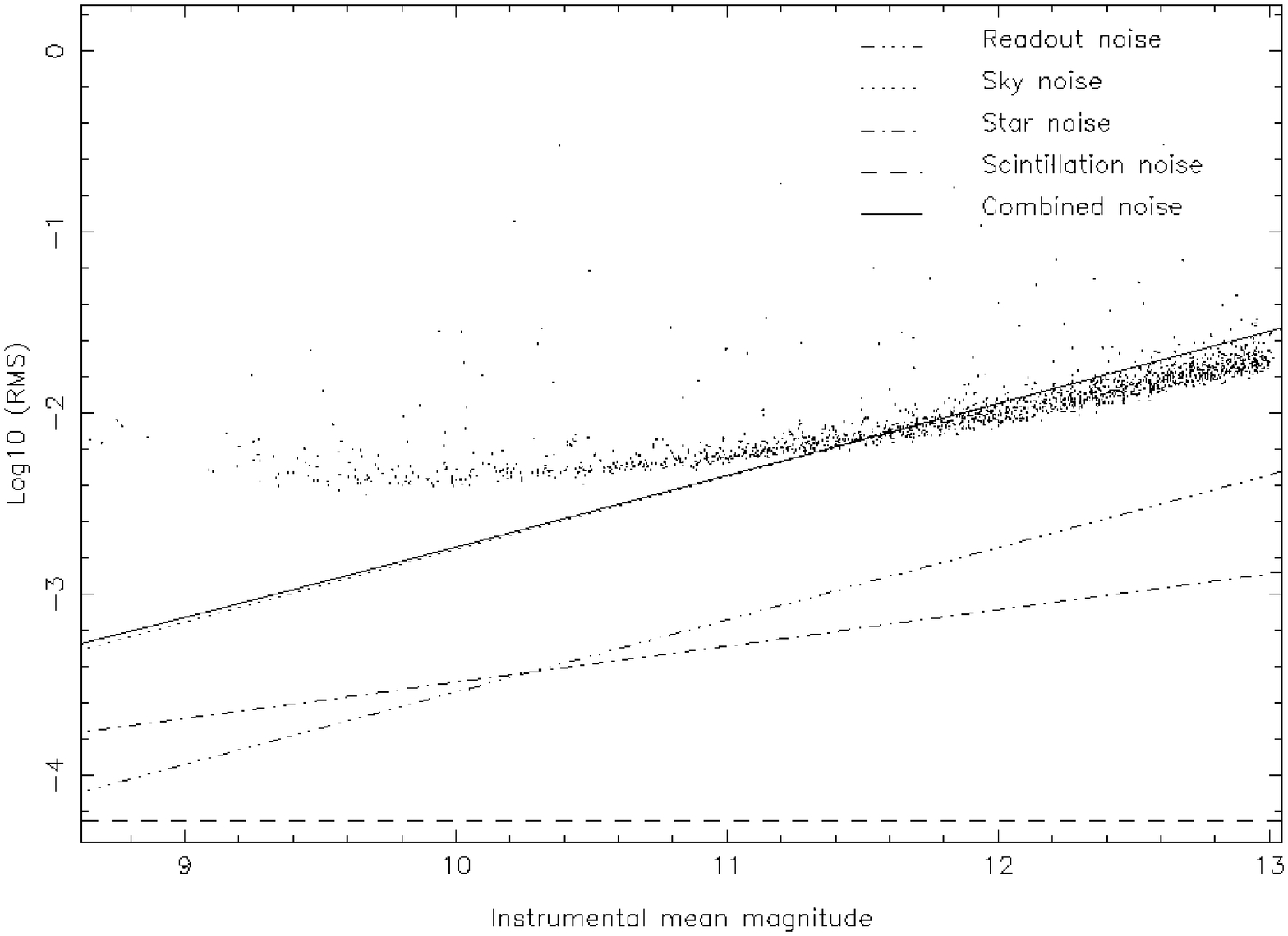}
\end{figure}

\clearpage

\begin{figure}
  \centering
     \includegraphics[width=100mm, height=100mm, angle=-90]{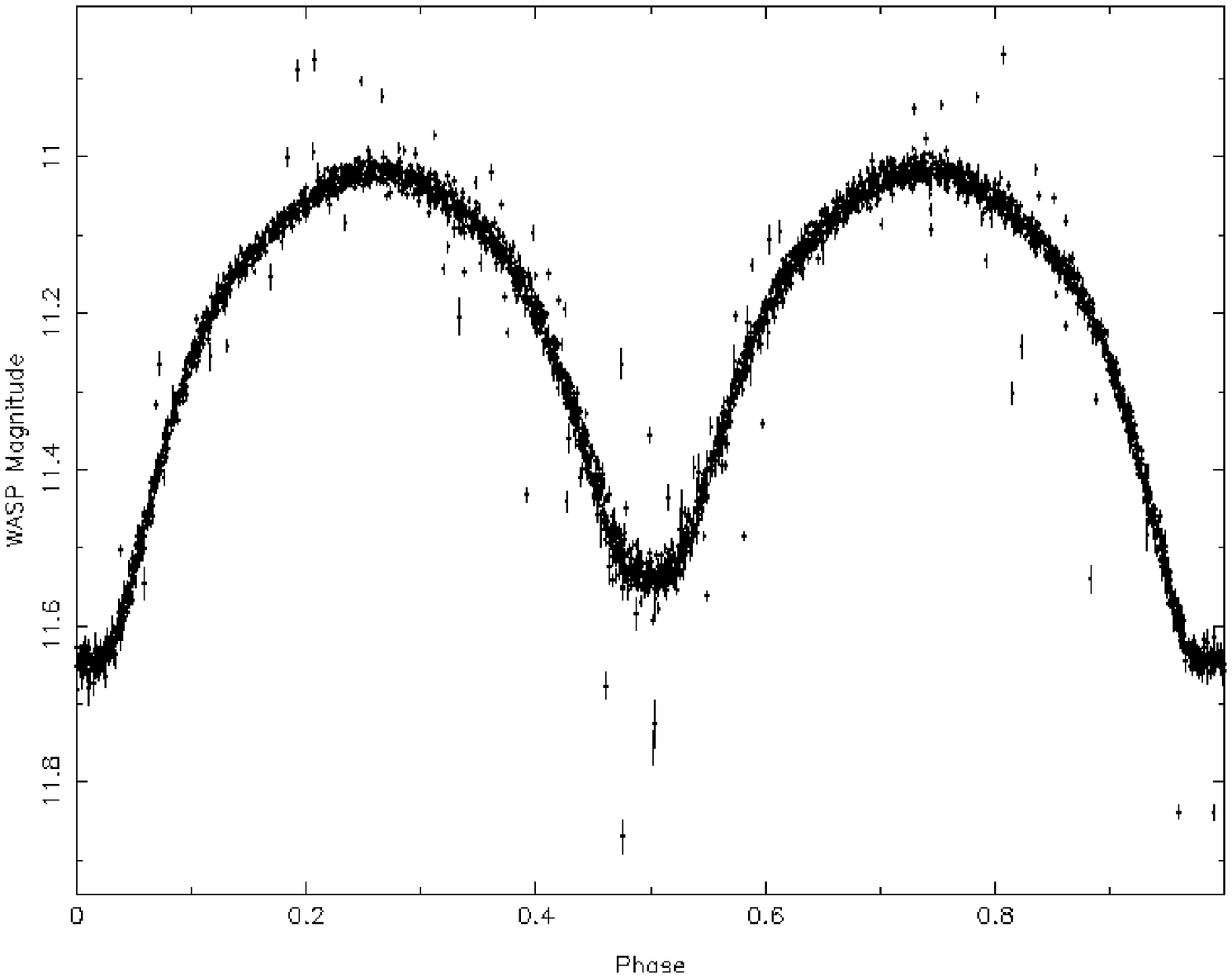}
\end{figure}

\clearpage

\begin{figure}
  \centering
     \includegraphics[width=100mm, angle=-90]{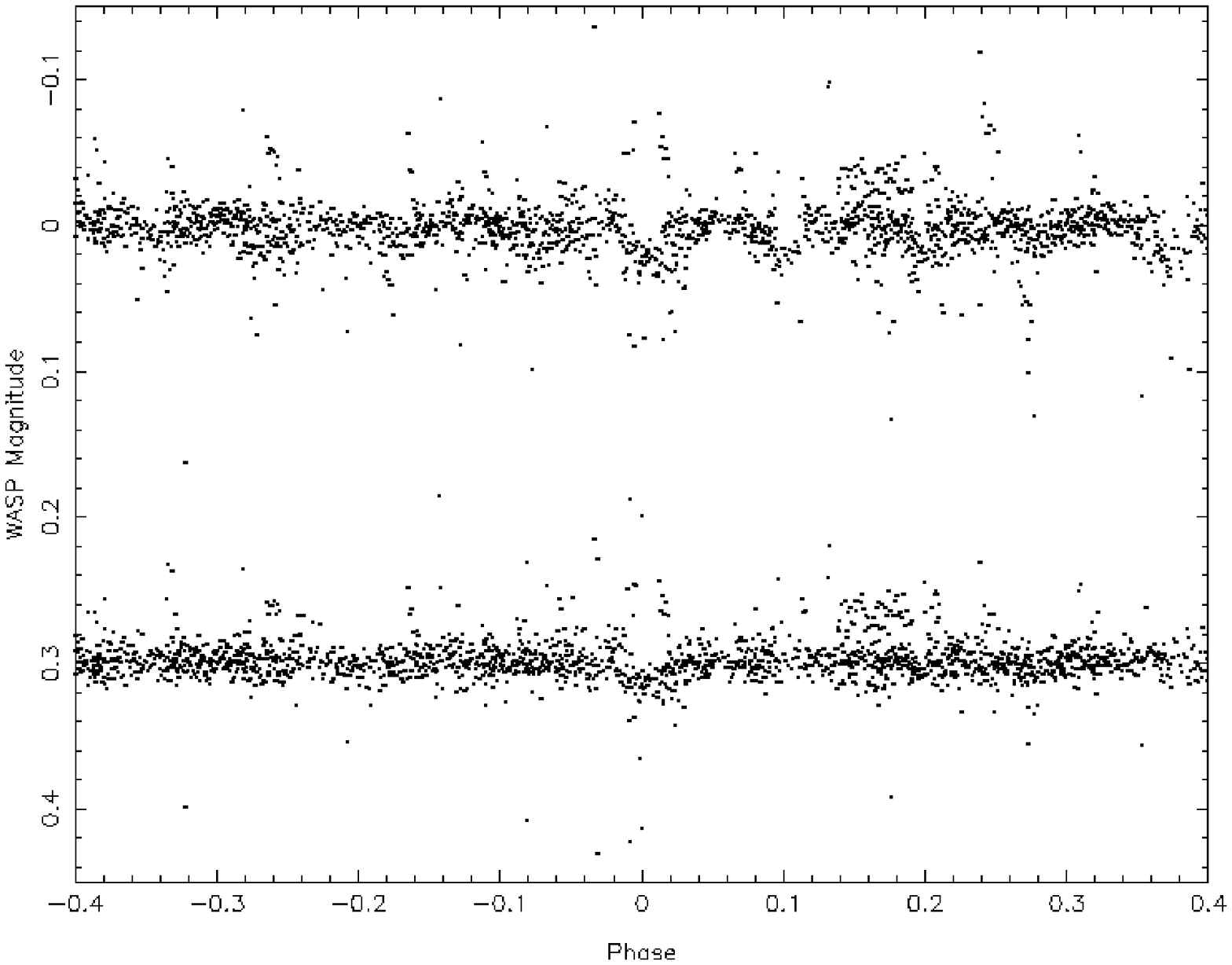}
\end{figure}

\end{document}